# Fairness Deconstructed: A Sociotechnical View of 'Fair' Algorithms in Criminal Justice

Rajiv Movva

Massachusetts Institute of Technology

rmovva@mit.edu

**Abstract**

Early studies of risk assessment algorithms used in criminal justice revealed widespread racial biases. In response, machine learning researchers have developed methods for fairness, many of which rely on equalizing empirical metrics across protected attributes. Here, I recall socio-technical perspectives to delineate the significant gap between fairness in theory and practice, focusing on criminal justice. I (1) illustrate how social context can undermine analyses that are restricted to an AI system's outputs, and (2) argue that much of the fair ML literature fails to account for epistemological issues with underlying crime data. Instead of building AI that reifies power imbalances, like risk assessment algorithms, I ask whether data science can be used to understand the root causes of structural marginalization.

## 1 Introduction

The United States criminal justice system has long been criticized for its discriminatory treatment of marginalized groups. At every stage in the pipeline, African Americans are multiple factors more likely to experience severe punishment than white Americans[1]. In this unequal chain of events, bail represents the first link: traditionally, a judge will set a price that the defendant must post to avoid being jailed before their hearing. Evidently, monetary bail inherently privileges wealthier defendants, without considering their risk to the public or likelihood to adhere to the judicial process. This explicit misprioritization has disproportionately affected the poor, multiplying existing race and class biases so as to predestine certain individuals to lives of imprisonment[2].

Citing violations to due process and equal protection, civil liberties organizations and other advocates have long fought for bail reform. Such movements have gained varying degrees of traction across America[3]. In many states, judges must now account for a defendant's financial status, likelihood of failing to appear, and immediate community risk in determining bail. In other states like New York and New Jersey, recent legislation has abolished cash bail and prevented pretrial jailing for non-violent crimes. However, while cash bail discriminates explicitly, any bail system that relies on judicial discretion leaves room for the effects of implicit bias.

---

[1] "Report to the United Nations on Racial Disparities in the U.S. Criminal Justice System," The Sentencing Project, accessed April 19, 2020, https://www.sentencingproject.org/publications/un-report-on-racial-disparities/.

[2] Marc Mauer, "Race, Class, and the Development of Criminal Justice Policy," *Review of Policy Research* 21, no. 1 (2004): 79–92, https://doi.org/10.1111/j.1541-1338.2004.00059.x.

[3] Alexa Van Brunt and Locke Bowman, "Toward a Just Model of Pretrial Release: A History of Bail Reform and a Prescription for What's Next," *Journal of Criminal Law and Criminology* 108, no. 4 (January 1, 2018): 701.

Alongside modern bail reform movements, the last decade has seen exponential growth in the volume and success of work on artificial intelligence (AI). On many reasoning tasks, AI has justifiably built reputation as more accurate, more efficient, and more consistent than human judgment. Such data-driven decision-making has also made its way into pretrial assessment, where supposedly objective algorithms are seen as a potential route to mitigating unequal treatment. In this setting, a defendant's risk is algorithmically quantified at their arraignment, and judges consider this score in deciding on release, bail, or detainment.

However, a now well-established body of work has challenged the normative assumption that algorithms are unbiased. In 2017, a seminal study from Buolamwini and Gebru showed that facial recognition algorithms are woefully inaccurate at detecting Black female faces compared to white males[4]. A recent *Science* paper demonstrated that a widely-deployed algorithm in hospitals to determine a patient's healthcare needs allocated fewer resources to Black patients than white patients in equally severe conditions[5]. In pretrial risk prediction, a ProPublica analysis revealed disparate algorithmic treatment of Black and white people: while the overall accuracy was the same for both groups, the AI system skewed towards predicting Black people as more risky than the data suggested, and white people as less.[6] Academics have largely internalized the danger that, if gone unchecked, algorithms will reproduce biases of their underlying datasets.

In response, researchers involved with the creation and deployment of these algorithms have proposed a number of adjustments to design fair, unbiased AI systems. On the technical side, theoretical computer scientists have pursued algorithmic fairness by defining mathematical criteria modeled after legal concepts like "equal protection" and "disparate impact"[7,8]. Meanwhile, computational social scientists have focused on identifying best practices for the design and use of algorithmic systems that tend to center around more representative dataset curation and increased transparency of the underlying AI models[9,10]. While the field has not yet converged on any single metric or set of practices, many seem to believe that algorithmic fairness is contingent on finding the correct procedure[11].

Here, I argue that existing notions of fairness are insufficient to rescue the damages caused by pretrial risk assessment algorithms. By conceiving of fairness as a function of an AI model's inputs and outputs, we ignore the dataset's epistemological assumptions and the algorithm's social context. Rather than blindly optimizing these tools, we should challenge their underlying premises and question their potential benefits. In doing so, I claim that AI for pretrial risk prediction serves to reinforce and qualify discriminatory treatment via objective validation, depoliticizing the issue. I close with a call to reframe data science in criminal justice as a study of systems rather than individuals. Instead of attempting to quantify a person's criminality, algorithms can be recast as a method of studying how communities are criminalized in discriminatory ways.

[4] Joy Buolamwini and Timnit Gebru, "Gender Shades: Intersectional Accuracy Disparities in Commercial Gender Classification," in *Conference on Fairness, Accountability and Transparency*, 2018, 77–91, http://proceedings.mlr.press/v81/buolamwini18a.html.

[5] Ziad Obermeyer et al., "Dissecting Racial Bias in an Algorithm Used to Manage the Health of Populations," *Science* 366, no. 6464 (October 25, 2019): 447–53, https://doi.org/10.1126/science.aax2342.

[6] Jeff Larson Julia Angwin, "Machine Bias," text/html, ProPublica (ProPublica, May 23, 2016), https://www.propublica.org/article/machine-bias-risk-assessments-in-criminal-sentencing.

[7] Sam Corbett-Davies et al., "Algorithmic Decision Making and the Cost of Fairness," in *Proceedings of the 23rd ACM SIGKDD International Conference on Knowledge Discovery and Data Mining*, KDD '17 (Halifax, NS, Canada: Association for Computing Machinery, 2017), 797–806, https://doi.org/10.1145/3097983.3098095.

[8] Michael Feldman et al., "Certifying and Removing Disparate Impact," in *Proceedings of the 21th ACM SIGKDD International Conference on Knowledge Discovery and Data Mining*, KDD '15 (Sydney, NSW, Australia: Association for Computing Machinery, 2015), 259–268, https://doi.org/10.1145/2783258.2783311.

[9] Timnit Gebru et al., "Datasheets for Datasets," *ArXiv:1803.09010 [Cs]*, March 19, 2020, http://arxiv.org/abs/1803.09010.

[10] Margaret Mitchell et al., "Model Cards for Model Reporting," *Proceedings of the Conference on Fairness, Accountability, and Transparency - FAT\* '19*, 2019, 220–29, https://doi.org/10.1145/3287560.3287596.

[11] Sam Corbett-Davies and Sharad Goel, "The Measure and Mismeasure of Fairness: A Critical Review of Fair Machine Learning," *ArXiv:1808.00023 [Cs]*, August 14, 2018, http://arxiv.org/abs/1808.00023.

## 2    The Promise of AI in Bail Reform

### 2.1    The Call for Fair Pretrial Assessment

A growing body of recent literature studying the criminal justice pipeline has identified widespread discrimination in pretrial practices[12,13,14,15]. Pierson et al. find that police are more likely to stop, ticket, and arrest Black drivers, and that the bar for searching them is much lower than for searching white drivers. Arnold et al. study bias in bail decisions, finding that white defendants who are not detained are significantly more likely to engage in pretrial misconduct than Black defendants who are not detained (suggesting that white defendants are released before trial much more readily). This double standard of pretrial imprisonment has long-term effects: Dobbie et al. and Leslie and Pope find that detainment increases conviction and recidivism rates, and reduces chances of future employment.

Perhaps afraid to further ignite tensions between the justice complex and marginalized groups, many researchers point to unavoidable implicit bias as the source of unequal treatment. Further, psychology research on cognitive biases supports this notion, with several definitions that rationalize a person's tendency to rely on stereotypes and make snap judgments. Importantly, these types of rhetoric frame bias as a failure of the individual rather than the system, concealing structural marginalization under the guise of being human and making natural mistakes. This framing also lends itself naturally to the inclusion of AI systems, since they are the canonical emotionless decision-makers that can overcome cognitive lapses[16].

With discrimination left unchallenged as merely a byproduct of human error, data becomes the neutral, objective truth. So long as we can get a machine to effectively model it, we've escaped the dangers of human biases. Across the country, this line of thinking has spawned dozens of pretrial risk assessment algorithms in order to objectively ground the opinions of fickle-minded judges. These algorithms use a proprietary set of features about a defendant to try and infer their risk. The COMPAS algorithm, for example, predicts on a scale of 1 to 10 how likely a defendant is to commit another crime in the future.[17] While its parent company Northpointe is as murky as possible about specifics, the algorithm is known to use a wide range of inputs: demographic indicators, personality questions, family background, and more[18]. Race is not directly considered, but can be easily inferred from correlated variables that the model does use.

---

[12] Emma Pierson et al., "A Large-Scale Analysis of Racial Disparities in Police Stops across the United States," *ArXiv:1706.05678 [Stat]*, June 18, 2017, http://arxiv.org/abs/1706.05678.

[13] David Arnold, Will Dobbie, and Crystal S. Yang, "Racial Bias in Bail Decisions," *The Quarterly Journal of Economics* 133, no. 4 (November 1, 2018): 1885–1932, https://doi.org/10.1093/qje/qjy012.

[14] Will Dobbie, Jacob Goldin, and Crystal S. Yang, "The Effects of Pretrial Detention on Conviction, Future Crime, and Employment: Evidence from Randomly Assigned Judges," *American Economic Review* 108, no. 2 (February 2018): 201–40, https://doi.org/10.1257/aer.20161503.

[15] Emily Leslie and Nolan G. Pope, "The Unintended Impact of Pretrial Detention on Case Outcomes: Evidence from New York City Arraignments," *The Journal of Law and Economics* 60, no. 3 (August 1, 2017): 529–57, https://doi.org/10.1086/695285.

[16] Matthew DeMichele et al., "The Intuitive-Override Model: Nudging Judges Toward Pretrial Risk Assessment Instruments," SSRN Scholarly Paper (Rochester, NY: Social Science Research Network, April 25, 2018), https://doi.org/10.2139/ssrn.3168500.

[17] Julia Angwin, "Machine Bias."

A selling point of AI systems like COMPAS is their ability to lock down predictively salient variables, and ignore others. This behavior is contrasted with the purported subjectivity of human judges. However, while judges must provide at least some nominal rationale for their decisions, the algorithm's output is not forced to be explained: black-box algorithms are accepted, so long as they achieve a satisfactory accuracy score. This romanticization of accuracy and other quantifiable metrics represents a major shortcoming of supposedly fair algorithms, an idea I explore in section 3.2.

For all they advertise, currently deployed pretrial risk assessment algorithms fail egregiously. While scrutinizing these systems can be very difficult because of the barriers to accessing relevant statistics, a number of exposés of algorithms in the justice system have been published[19,20,21]. For example, the previously mentioned ProPublica study scrutinizes the use of COMPAS in Florida's Broward County. While Northpointe claims that COMPAS is equally accurate for white and Black people, when the system does make errors, it does so differently for different groups. In particular, Black people are more often falsely predicted to commit another crime, while white people that actually do end up committing another crime are often predicted innocent. This example suggests how theoretically correct statements about a model's accuracy hides specifics about its practical effects. Thus, enter the community of fairness-conscious computer scientists attempting to find a better set of metrics or practices to reduce bias.

## 2.2   Theoretical Notions of Fairness

As studies have increasingly demonstrated that algorithms can fail to handle diverse input data, an ethical AI subcommunity has emerged in order to study and fix these systems. Much of this work is devoted to identifying theoretical metrics that best expose how algorithms practically make mistakes. Legal principles like equal protection and consideration are considered in terms of mathematical formulas, with researchers arguing that previous discussions lack precision and often substitute "rhetoric ... for careful analysis"[22].

The field has far from agreed on a single fairness formula that best describes all situations. Many different metrics have been proposed, some of which are inspired by particular legal ideas (such as "disparate impact" from the EEOC[23]), and authors make cases for why their metric is best using a specific validation scheme. Among others, Berk et al. have begun to perform meta-studies comparing different criteria, and find that optimizing one fairness formula often worsens another[24]. Another recurring theme is that fairness comes at a cost: for a system to be more fair, it must be less accurate, and vice-versa[25]. Hence, the research community has moved towards a defining set of trade-offs that must be balanced in order to inform policy. It is taken as fact that no tool will fix systemic racial or gender inequities in the legal system. At best, scientists can delineate a trade-off curve that formalizes ideas of fairness and accuracy and makes their relationship more transparent[26]. It is then the job of policymakers and the public to decide which specific point on the algorithmic curve is most appropriate for a given social context. In the realm of criminal justice, Corbett-Davies et al. explain that there is difficulty in such a decision because one must resolve seemingly unavoidable "tension between improving public safety and satisfying … algorithmic fairness". Importantly, this discourse makes the ideological leap of equating an algorithm's accuracy to its ability to effect public safety, an idea that I challenge in section 3.2.

[18] Electronic Privacy Information Center, "EPIC - Algorithms in the Criminal Justice System: Pre-Trial Risk Assessment Tools," accessed April 6, 2020, https://epic.org/algorithmic-transparency/crim-justice/.
[19] Julia Angwin, "Machine Bias."
[20] Cade Metz and Adam Satariano, "An Algorithm That Grants Freedom, or Takes It Away," *The New York Times*, February 6, 2020, sec. Technology, https://www.nytimes.com/2020/02/06/technology/predictive-algorithms-crime.html.
[21] Nick Wingfield, "Amazon Pushes Facial Recognition to Police. Critics See Surveillance Risk.," *The New York Times*, May 22, 2018, sec. Technology, https://www.nytimes.com/2018/05/22/technology/amazon-facial-recognition.html.
[22] Richard Berk et al., "Fairness in Criminal Justice Risk Assessments: The State of the Art," *ArXiv:1703.09207 [Stat]*, May 27, 2017, http://arxiv.org/abs/1703.09207.
[23] Michael Feldman et al., "Certifying and Removing Disparate Impact," in *Proceedings of the 21th ACM SIGKDD International Conference on Knowledge Discovery and Data Mining*, KDD '15 (Sydney, NSW, Australia: Association for Computing Machinery, 2015), 259–268, https://doi.org/10.1145/2783258.2783311.
[24] Berk et al., "Fairness in Criminal Justice Risk Assessments."

# 3     Practical Pitfalls of Fairness

Much of the current work on ethical AI implicitly relies on two assumptions. First, several papers discuss bias as if the discriminatory effects of using an algorithmic system can be entirely modeled *in silico*[27]. For example, quantifications of algorithmic bias are extrapolated to reflect the true extent of bias that would exist in a pretrial risk assessment setting. Second, historical arrest data are read as rigid attributes of individuals, rather than as numerical descriptions of how police structures interact with various groups. In section 3.1, I subvert these assumptions and argue that existing work on fairness ceases to be relevant when we leave the "algorithmic frame". In section 3.2, I discuss the epistemological assumptions regarding crime data that are used to justify risk prediction algorithms. Left unquestioned, these assumptions can cause algorithms to legitimize oppressive forms of 'justice' and exacerbate incarceration of minorities.

## 3.1     Social Context Undermines Theoretical Fairness

Building a machine learning model starts with finding a dataset and mapping that data to inputs and outputs. Such design choices are typically made at the start of a project, and tend to remain fixed. Generally, fairness researchers abide by the same structure: for example, the dataset will consist of crimes committed in Broward County, inputs will consist of various human attributes (à la COMPAS), and the output will be recidivism risk[28]. Having defined the scope of their prediction system, researchers enter the "algorithmic frame", and thus any discussion of accuracy or fairness exists in this specific context.

These narrow bounds fail to capture the full complexity of fairness and its sociocultural meanings. In the context of criminal justice, fairness refers to the expectation that the legal process does not favor any individual a priori. However, as soon as a researcher decides on particular algorithmic inputs and outputs, their system is divorced from the broader social context in which it is applied. Thus, there exists a gaping discrepancy between our human understanding of fairness and mathematical formulas for fairness.

---

[25] Corbett-Davies et al., "Algorithmic Decision Making and the Cost of Fairness."
[26] Berk et al., "Fairness in Criminal Justice Risk Assessments."
[27] Andrew D. Selbst et al., "Fairness and Abstraction in Sociotechnical Systems," SSRN Scholarly Paper (Rochester, NY: Social Science Research Network, August 23, 2018), https://papers.ssrn.com/abstract=3265913.
[28] Corbett-Davies et al., "Algorithmic Decision Making and the Cost of Fairness."

Selbst et al. term this conflation of algorithmic fairness with true fairness the 'framing trap'[29]. In contrast with the narrow algorithmic frame, they define a *sociotechnical frame* as one that explicitly acknowledges AI's role as a component in a larger process. In the sociotechnical frame, arraignments are resolved with a judge's decision of release on personal recognizance, release with money bail, or detainment. The process by which this decision is made and the outcomes across individuals are what determine our evaluation of fairness. I highlight three ways in which algorithmic fairness fails to align with this vision of fairness.

First, most pretrial algorithms claim to measure *risk*, which can refer to a broad range of outputs. COMPAS predicts any type of future recidivism[30]. More tailored algorithms predict failure to appear at a hearing, or the chance of the defendant committing an additional crime while released[31]. Even if one can demonstrate that an algorithm predicts these measures fairly across groups, there is still a gap between the prediction and the arraignment decision. Risk is merely a proxy for the ultimate outcome. Hence, claims about equitable algorithmic behavior don't extend to fairness of the pretrial process.

Second, algorithmic definitions of fairness wash out the sociocultural nuance of the term. These metrics treat the algorithm as a black-box: the machine learning system will compute predicted outputs for individuals from a testing set that it did not see during training. These inputs and outputs alone determine whether or not a system is fair, irrespective of the model's underlying reasoning. While model explainability is a growing research direction, it is not directly accounted for in any algorithmic fairness formulas[32]. In contrast, our human understanding of fairness revolves around procedure: how an outcome is achieved, not the outcome itself, is what matters. As Selbst et al. point out, this understanding of procedural fairness is also legally codified *e.g.* in employment discrimination law[33]. In the context of bail, a system could be fair despite different outcomes across groups: for example, a fair system may release poorer individuals more on recognizance than money bail. Fairness has connotations that are not captured by outcome-focused criteria.

Third, there is evidence that the sociotechnical frame includes variance that is left unmodeled by algorithms. In particular, a study of pretrial risk assessment in Kentucky showed that judges do not always use predicted scores, and they use them in differing ways when they do[34]. Given enough data on particular judges, researchers have proposed algorithmically modeling individual behavior, but many judges also vary how they use the recommendations over time[35,36]. Fundamentally, judges have full discretion over the algorithm with no real accountability, easily creating scenarios where discrimination is maintained. For example, influenced by confirmation bias, a judge may choose to cite the algorithm as evidence only when it aligns with their opinion. Many fairness papers offer numerical guarantees on the maximum racial disparity of their algorithmic predictions. Not knowing when and how their scores are actually used severely undermines this work, as judges can easily reintroduce the bias researchers worked so hard to erase.

[29] Andrew D. Selbst et al., "Fairness and Abstraction in Sociotechnical Systems," SSRN Scholarly Paper (Rochester, NY: Social Science Research Network, August 23, 2018), https://papers.ssrn.com/abstract=3265913.
[30] Julia Angwin, "Machine Bias."
[31] Megan T. Stevenson, "Assessing Risk Assessment in Action," SSRN Scholarly Paper (Rochester, NY: Social Science Research Network, 2018), https://doi.org/10.2139/ssrn.3016088.
[32] Cynthia Rudin and Joanna Radin, "Why Are We Using Black Box Models in AI When We Don't Need To? A Lesson From An Explainable AI Competition," *Harvard Data Science Review* 1, no. 2 (November 1, 2019), https://doi.org/10.1162/99608f92.5a8a3a3d.
[33] Selbst et al., "Fairness and Abstraction in Sociotechnical Systems." At 6.
[34] Stevenson, "Assessing Risk Assessment in Action."
[35] Ibid.
[36] Chelsea Barabas et al., "Studying up: Reorienting the Study of Algorithmic Fairness around Issues of Power," in *Proceedings of the 2020 Conference on Fairness, Accountability, and Transparency*, FAT* '20 (Barcelona, Spain: Association for Computing Machinery, 2020), 167–176, https://doi.org/10.1145/3351095.3372859.

## 3.2    AI Misinterprets Its Data

The fair AI community has largely come to terms with the fact that no single metric will best model fairness[37]. Further, the aforementioned concerns surrounding the gaps between theoretical and practical fairness have started being broadly acknowledged[38,39,40]. Still, many researchers make the argument that perfect shouldn't be the enemy of the good, arguing that an algorithm should be used as long as it performs better than judges[41]. In this line of reasoning, algorithms and judges are compared via metrics like accuracy and bias.

However, measuring the correctness of a judge's decision is statistically impossible, because we cannot tell whether someone who was detained would've actually committed a crime or not. Researchers have developed estimation methods to try and overcome this issue[42,43]. Kleinberg et al. report that algorithmic decisions incarcerate fewer people with no increase in crime rates, and Jung et al. find that their system "significantly outperforms judges" in accuracy. While this research does acknowledge the limitations of its calculations, accuracy and bias are still seen as the primary measuring sticks to determine an algorithm's efficacy.

In using these metrics as the end-all-be-all, we fail to challenge the veracity of the crime datasets behind these algorithms. Pretrial assessment algorithms use arrest statistics as a proxy for an individual's riskiness – but what do these numbers really mean? As civil rights organizations correctly point out, "such data primarily document the behavior and decisions of police officers and prosecutors, rather than the individuals or groups that the data are claiming to describe"[44]. Decades of research have proven that groups are policed differently: drug laws and their enforcement, for example, result in vastly different treatment between Black and white Americans[45,46]. Science has helped rigorously surface these quantitative disparities;

---

[37] Corbett-Davies and Goel, "The Measure and Mismeasure of Fairness."
[38] Selbst et al., "Fairness and Abstraction in Sociotechnical Systems."
[39] Abigail Z. Jacobs and Hanna Wallach, "Measurement and Fairness," *ArXiv:1912.05511 [Cs]*, December 11, 2019, http://arxiv.org/abs/1912.05511.
[40] Kate Crawford and Ryan Calo, "There Is a Blind Spot in AI Research," *Nature News* 538, no. 7625 (October 20, 2016): 311, https://doi.org/10.1038/538311a.
[41] Sam Corbett-Davies, Sharad Goel, and Sandra González-Bailón, "Even Imperfect Algorithms Can Improve the Criminal Justice System," *The New York Times*, December 20, 2017, sec. The Upshot, https://www.nytimes.com/2017/12/20/upshot/algorithms-bail-criminal-justice-system.html.
[42] Jongbin Jung et al., "Simple Rules for Complex Decisions," *ArXiv:1702.04690 [Stat]*, April 2, 2017, http://arxiv.org/abs/1702.04690.
[43] Jon Kleinberg et al., "Human Decisions and Machine Predictions," *The Quarterly Journal of Economics* 133, no. 1 (February 1, 2018): 237–93, https://doi.org/10.1093/qje/qjx032.
[44] "More than 100 Civil Rights, Digital Justice, and Community-Based Organizations Raise Concerns About Pretrial Risk Assessment," The Leadership Conference on Civil and Human Rights, July 30, 2018, https://civilrights.org/2018/07/30/more-than-100-civil-rights-digital-justice-and-community-based-organizations-raise-concerns-about-pretrial-risk-assessment/.
[45] Ojmarrh Mitchell and Michael S. Caudy, "Examining Racial Disparities in Drug Arrests," *Justice Quarterly* 32, no. 2 (March 4, 2015): 288–313, https://doi.org/10.1080/07418825.2012.761721.
[46] Deborah J. Vagins and Jesselyn McCurdy, "Cracks in the System: 20 Years of the Unjust Federal Crack Cocaine Law," American Civil Liberties Union, accessed April 22, 2020, https://www.aclu.org/other/cracks-system-20-years-unjust-federal-crack-cocaine-law.

ironically, these socially constructed differences are now being marked as useful signal that helps us understand criminals.

Rather than highlighting the circular reasoning here, the latest rhetoric from some in the fair ML community plays into the fallacy. Established statistician Richard Berk argues that accurate algorithms *should* perform differently across groups, because groups have widely varying "base rates" of criminal activity[47]. These base rates – which typically refer to arrests – are uncritically portrayed as universal truths embedded in a group's existence. In this framing, Black people are inherently riskier than white people, and deserve to be algorithmically handled as such. Corbett-Davies et al. extend this discourse to argue that an accurate system has direct public safety benefits[48]. They bring up the theoretical example of an algorithm that could predict a defendant's chance of committing a *violent* crime on pretrial release, and argue that the fairness of such a system should be balanced against the lives it could save. However, the rate of pretrial violent crime is so low that an accurate system would always predict that a person is unlikely to commit one[49]. Instead, because datasets consider even minor crimes as criminal activity[50], overpoliced groups are assigned high risk scores, judges misinterpret these scores, and far more innocent people are detained than necessary.

In leaving datasets unchallenged and taking crime statistics at face value, researchers are only calcifying oppressive ideologies. Historical datasets are rife with racism, and AI justifies that racism as a pragmatic public safety measure. This type of data misinterpretation isn't new: in the 1990s, law enforcement used crime statistics to justify racial profiling. As more data were collected, it became apparent that Black and Latine people were disproportionately arrested and jailed. In response, these communities were policed even further, because the data clearly proved that they were most criminal[51]. As Chelsea Barabas points out, "AI is the most recent incarnation of this historical struggle over the interpretation of justice data"[52]. Much work on ethical AI has claimed to make algorithms fair without critically studying the practices underlying the datasets. This negligence, and the new sense of technical legitimacy that AI brings, only further obscures the deep-seated inequity present in our justice system.

## 4        Reimagining AI in Criminal Justice

So far, I have asserted that current modes of artificial intelligence are not an effective approach to bail reform. Research in fair machine learning has yet to lock down a definition of fairness that properly models the social context of the criminal justice system. I've argued that such a 'fair' algorithm is in fact impossible to create, because currently available data is incompatible with developing an unbiased system.

Noting many of these same points, social scientists have started to bring an abolitionist view to the field. Given an arbitrary problem involving inputs and outputs, a computer scientist's worldview tends to believe that an algorithmic solution is best and works off that assumption. Academics from other domains have started to temper this technological manifest destiny by questioning whether an algorithm really is the best approach[53,54]. Civil rights groups have recently put out a consensus statement against the use of pretrial risk assessment algorithms, calling for mitigation where already in use and lobbying against adoption elsewhere.[55] Indeed, abolitionist activism is critical in this early stage, before AI becomes ingrained in the system.

---

[47] Berk et al., "Fairness in Criminal Justice Risk Assessments."
[48] Corbett-Davies and Goel, "The Measure and Mismeasure of Fairness."
[49] Chelsea Barabas, "Beyond Bias: Re-Imagining the Terms of 'Ethical AI' in Criminal Law," SSRN Scholarly Paper (Rochester, NY: Social Science Research Network, April 25, 2019), https://doi.org/10.2139/ssrn.3377921.
[50] Kleinberg et al., "Human Decisions and Machine Predictions."
[51] David Harris, "The Reality of Racial Disparity in Criminal Justice: The Significance of Data Collection," *Law and Contemporary Problems* 66, no. 3 (July 1, 2003): 71–98.
[52] Barabas, "Beyond Bias."

Well, *can* data science positively impact the criminal justice system? So far, algorithms have focused almost exclusively on studying the marginalized. Risk prediction brings additional scrutiny to those that are targeted most by the carceral state. Data science "looks down", largely because "the people that have access to data and to the technical skills to work with it are those that have the most stake in reproducing the status quo"[56]. When the AI community is full of privilege, it's easy for statements like those about inherent base rates of marginalized people's dangerousness to go unchecked. And the criminal justice system has a direct stake in biased algorithms, since those systems help reify their position of power.

It's time to turn the microscope away from the marginalized. Instead of ramping up surveillance of those without power, we can use data science to examine the structures that create oppression. In Sam Corbett-Davies' New York Times op-ed, he cites a statistic that the strictest judges in NYC are more than twice as likely to demand bail than the most lenient ones[57]. He uses this point of judges' capriciousness to call for pretrial risk algorithms; instead, why don't we perform more systematic analysis on how judges act? In their call to "study up" and examine the individuals *with* power in criminal justice, Chelsea Barabas et al. undertake such an attempt[58]. Importantly, they find that accessing detailed data on the decisions that a judge makes is incredibly difficult. There are significant barriers explicitly created to maintain data privacy for individuals in power; the marginalized are afforded no such protection. Using what judge-specific data they could find, the authors built a so-called judicial risk assessment algorithm. The algorithm quantifies a judge's tendency to violate the Constitution "by imposing unaffordable bail without due process"[59]. While the work was meant more as a thought exercise than a pragmatic tool, it shows one of the reciprocal ways we can use AI while shifting focus from the oppressed to the oppressor.

Algorithmic tools must be more carefully challenged. Many applications of AI may serve to reinforce bias, and branding a system as fair via mathematical formalisms is not the solution. Rather, one must carefully consider the sociocultural context in which an algorithm is applied, and use this context to critically interrogate the algorithm's epistemological assumptions. In criminal justice, I believe that taking these steps renders pretrial risk systems unethical. Data science should instead be used to study the structures underpinning the systemic criminalization of minorities by the carceral state.

---

[53] Selbst et al., "Fairness and Abstraction in Sociotechnical Systems."
[54] Barabas et al., "Studying Up."
[55] "More than 100 Civil Rights, Digital Justice, and Community-Based Organizations Raise Concerns About Pretrial Risk Assessment."
[56] Catherine D'Ignazio and Lauren Klein, "Chapter Seven: The Power Chapter," in *Data Feminism* (PubPub, 2018), https://bookbook.pubpub.org/pub/7ruegkt6.
[57] Corbett-Davies, Goel, and González-Bailón, "Even Imperfect Algorithms Can Improve the Criminal Justice System."
[58] Barabas et al., "Studying Up."
[59] Ibid.

## Works Cited

Arnold, David, Will Dobbie, and Crystal S. Yang. "Racial Bias in Bail Decisions." *The Quarterly Journal of Economics* 133, no. 4 (November 1, 2018): 1885–1932. https://doi.org/10.1093/qje/qjy012.

Barabas, Chelsea. "Beyond Bias: Re-Imagining the Terms of 'Ethical AI' in Criminal Law." SSRN Scholarly Paper. Rochester, NY: Social Science Research Network, April 25, 2019. https://doi.org/10.2139/ssrn.3377921.

Barabas, Chelsea, Colin Doyle, JB Rubinovitz, and Karthik Dinakar. "Studying up: Reorienting the Study of Algorithmic Fairness around Issues of Power." In *Proceedings of the 2020 Conference on Fairness, Accountability, and Transparency*, 167–176. FAT* '20. Barcelona, Spain: Association for Computing Machinery, 2020. https://doi.org/10.1145/3351095.3372859.

Berk, Richard. *Richard Berk - Using ML in Criminal Justice Risk Assessments - The Frontiers of Machine Learning*. Accessed April 19, 2020. https://www.youtube.com/watch?v=gdEPPRhNu34.

Berk, Richard, Hoda Heidari, Shahin Jabbari, Michael Kearns, and Aaron Roth. "Fairness in Criminal Justice Risk Assessments: The State of the Art." *ArXiv:1703.09207 [Stat]*, May 27, 2017. http://arxiv.org/abs/1703.09207.

Buolamwini, Joy, and Timnit Gebru. "Gender Shades: Intersectional Accuracy Disparities in Commercial Gender Classification." In *Conference on Fairness, Accountability and Transparency*, 77–91, 2018. http://proceedings.mlr.press/v81/buolamwini18a.html.

Caliskan, Aylin, Joanna J. Bryson, and Arvind Narayanan. "Semantics Derived Automatically from Language Corpora Contain Human-like Biases." *Science* 356, no. 6334 (April 14, 2017): 183–86. https://doi.org/10.1126/science.aal4230.

Caruana, Rich, Yin Lou, Johannes Gehrke, Paul Koch, Marc Sturm, and Noemie Elhadad. "Intelligible Models for HealthCare: Predicting Pneumonia Risk and Hospital 30-Day Readmission." In *Proceedings of the 21th ACM SIGKDD International Conference on Knowledge Discovery and Data Mining - KDD '15*, 1721–30. Sydney, NSW, Australia: ACM Press, 2015. https://doi.org/10.1145/2783258.2788613.

Center, Electronic Privacy Information. "EPIC - Algorithms in the Criminal Justice System: Pre-Trial Risk Assessment Tools." Accessed April 6, 2020. https://epic.org/algorithmic-transparency/crim-justice/.

Corbett-Davies, Sam, and Sharad Goel. "The Measure and Mismeasure of Fairness: A Critical Review of Fair Machine Learning." *ArXiv:1808.00023 [Cs]*, August 14, 2018. http://arxiv.org/abs/1808.00023.

Corbett-Davies, Sam, Sharad Goel, and Sandra González-Bailón. "Even Imperfect Algorithms Can Improve the Criminal Justice System." *The New York Times*, December 20, 2017, sec. The Upshot. https://www.nytimes.com/2017/12/20/upshot/algorithms-bail-criminal-justice-system.html.

Corbett-Davies, Sam, Emma Pierson, Avi Feller, Sharad Goel, and Aziz Huq. "Algorithmic Decision Making and the Cost of Fairness." In *Proceedings of the 23rd ACM SIGKDD International Conference on Knowledge Discovery and Data Mining*, 797–806. KDD '17. Halifax, NS, Canada: Association for Computing Machinery, 2017. https://doi.org/10.1145/3097983.3098095.

Crawford, Kate, and Ryan Calo. "There Is a Blind Spot in AI Research." *Nature News* 538, no. 7625 (October 20, 2016): 311. https://doi.org/10.1038/538311a.

Stanford Graduate School of Business. "Deep Neural Networks Are More Accurate Than Humans at Detecting Sexual Orientation From Facial Images." Accessed April 6, 2020. https://www.gsb.stanford.edu/faculty-research/publications/deep-neural-networks-are-more-accurate-humans-detecting-sexual.

DeMichele, Matthew, Megan Comfort, Shilpi Misra, Kelle Barrick, and Peter Baumgartner. "The Intuitive-Override Model: Nudging Judges Toward Pretrial Risk Assessment Instruments." SSRN Scholarly Paper. Rochester, NY: Social Science Research Network, April 25, 2018. https://doi.org/10.2139/ssrn.3168500.

D'Ignazio, Catherine, and Lauren Klein. "Chapter Seven: The Power Chapter." In *Data Feminism*. PubPub, 2018. https://bookbook.pubpub.org/pub/7ruegkt6.

Dobbie, Will, Jacob Goldin, and Crystal S. Yang. "The Effects of Pretrial Detention on Conviction, Future Crime, and Employment: Evidence from Randomly Assigned Judges." *American Economic Review* 108, no. 2 (February 2018): 201–40. https://doi.org/10.1257/aer.20161503.

American Civil Liberties Union. "Fair Sentencing Act." Accessed April 23, 2020. https://www.aclu.org/issues/criminal-law-reform/drug-law-reform/fair-sentencing-act.

Feldman, Michael, Sorelle A. Friedler, John Moeller, Carlos Scheidegger, and Suresh Venkatasubramanian. "Certifying and Removing Disparate Impact." In *Proceedings of the 21th ACM SIGKDD International Conference on Knowledge Discovery and Data Mining*, 259–268. KDD '15. Sydney, NSW, Australia: Association for Computing Machinery, 2015. https://doi.org/10.1145/2783258.2783311.

Gebru, Timnit, Jamie Morgenstern, Briana Vecchione, Jennifer Wortman Vaughan, Hanna Wallach, Hal Daumé III, and Kate Crawford. "Datasheets for Datasets." *ArXiv:1803.09010 [Cs]*, March 19, 2020. http://arxiv.org/abs/1803.09010.

Goel, Sharad, Ravi Shroff, Jennifer L. Skeem, and Christopher Slobogin. "The Accuracy, Equity, and Jurisprudence of Criminal Risk Assessment." SSRN Scholarly Paper. Rochester, NY: Social Science Research Network, December 26, 2018. https://doi.org/10.2139/ssrn.3306723.

Goncalves, Felipe, and Steven Mello. "A Few Bad Apples? Racial Bias in Policing." *Working Papers*. Working Papers. Princeton University, Department of Economics, Industrial Relations Section., March 2017. https://ideas.repec.org/p/pri/indrel/608.html.

Grother, Patrick J., Mei L. Ngan, and Kayee K. Hanaoka. "Face Recognition Vendor Test Part 3: Demographic Effects," December 19, 2019. https://www.nist.gov/publications/face-recognition-vendor-test-part-3-demographic-effects.

Harris, David. "The Reality of Racial Disparity in Criminal Justice: The Significance of Data Collection." *Law and Contemporary Problems* 66, no. 3 (July 1, 2003): 71–98.

Heilweil, Rebecca. "Why We Don't Know as Much as We Should about Police Surveillance Technology." Vox, February 5, 2020. https://www.vox.com/recode/2020/2/5/21120404/police-departments-artificial-intelligence-public-records.

Jacobs, Abigail Z., and Hanna Wallach. "Measurement and Fairness." *ArXiv:1912.05511 [Cs]*, December 11, 2019. http://arxiv.org/abs/1912.05511.

Jo, Eun Seo, and Timnit Gebru. "Lessons from Archives: Strategies for Collecting Sociocultural Data in Machine Learning." *ArXiv:1912.10389 [Cs]*, December 22, 2019. https://doi.org/10.1145/3351095.3372829.

Joh, Elizabeth E. "Automated Seizures: Police Stops of Self-Driving Cars." SSRN Scholarly Paper. Rochester, NY: Social Science Research Network, March 20, 2019. https://doi.org/10.2139/ssrn.3354800.

Julia Angwin, Jeff Larson. "Machine Bias." Text/html. ProPublica. ProPublica, May 23, 2016. Https://www.propublica.org/. https://www.propublica.org/article/machine-bias-risk-assessments-in-criminal-sentencing.

Jung, Jongbin, Connor Concannon, Ravi Shroff, Sharad Goel, and Daniel G. Goldstein. "Simple Rules for Complex Decisions." *ArXiv:1702.04690 [Stat]*, April 2, 2017. http://arxiv.org/abs/1702.04690.

Kleinberg, Jon, Himabindu Lakkaraju, Jure Leskovec, Jens Ludwig, and Sendhil Mullainathan. "Human Decisions and Machine Predictions." *The Quarterly Journal of Economics* 133, no. 1 (February 1, 2018): 237–93. https://doi.org/10.1093/qje/qjx032.

Kohl, Rhiana. "BAIL IN THE UNITED STATES: A BRIEF REVIEW OF THE LITERATURE," n.d., 4.

Leslie, Emily, and Nolan G. Pope. "The Unintended Impact of Pretrial Detention on Case Outcomes: Evidence from New York City Arraignments." *The Journal of Law and Economics* 60, no. 3 (August 1, 2017): 529–57. https://doi.org/10.1086/695285.

Lohr, Steve. "Facial Recognition Is Accurate, If You're a White Guy." *The New York Times*, February 9, 2018, sec. Technology. https://www.nytimes.com/2018/02/09/technology/facial-recognition-race-artificial-intelligence.html.

Mauer, Marc. "Race, Class, and the Development of Criminal Justice Policy." *Review of Policy Research* 21, no. 1 (2004): 79–92. https://doi.org/10.1111/j.1541-1338.2004.00059.x.

Metz, Cade, and Adam Satariano. "An Algorithm That Grants Freedom, or Takes It Away." *The New York Times*, February 6, 2020, sec. Technology. https://www.nytimes.com/2020/02/06/technology/predictive-algorithms-crime.html.

Mitchell, Margaret, Simone Wu, Andrew Zaldivar, Parker Barnes, Lucy Vasserman, Ben Hutchinson, Elena Spitzer, Inioluwa Deborah Raji, and Timnit Gebru. "Model Cards for Model Reporting." *Proceedings of the Conference on Fairness, Accountability, and Transparency - FAT\* '19*, 2019, 220–29. https://doi.org/10.1145/3287560.3287596.

Mitchell, Ojmarrh, and Michael S. Caudy. "Examining Racial Disparities in Drug Arrests." *Justice Quarterly* 32, no. 2 (March 4, 2015): 288–313. https://doi.org/10.1080/07418825.2012.761721.

The Leadership Conference on Civil and Human Rights. "More than 100 Civil Rights, Digital Justice, and Community-Based Organizations Raise Concerns About Pretrial Risk Assessment," July 30, 2018. https://civilrights.org/2018/07/30/more-than-100-civil-rights-digital-justice-and-community-based-organizations-raise-concerns-about-pretrial-risk-assessment/.

Mullainathan, Sendhil. "Biased Algorithms Are Easier to Fix Than Biased People." *The New York Times*, December 6, 2019, sec. Business. https://www.nytimes.com/2019/12/06/business/algorithm-bias-fix.html.

Murphy, Heather. "Why Stanford Researchers Tried to Create a 'Gaydar' Machine." *The New York Times*, October 9, 2017, sec. Science. https://www.nytimes.com/2017/10/09/science/stanford-sexual-orientation-study.html.

Obermeyer, Ziad, Brian Powers, Christine Vogeli, and Sendhil Mullainathan. "Dissecting Racial Bias in an Algorithm Used to Manage the Health of Populations." *Science* 366, no. 6464 (October 25, 2019): 447–53. https://doi.org/10.1126/science.aax2342.

Pierson, Emma, Camelia Simoiu, Jan Overgoor, Sam Corbett-Davies, Vignesh Ramachandran, Cheryl Phillips, and Sharad Goel. "A Large-Scale Analysis of Racial Disparities in Police Stops across the United States." *ArXiv:1706.05678 [Stat]*, June 18, 2017. http://arxiv.org/abs/1706.05678.

The Sentencing Project. "Report to the United Nations on Racial Disparities in the U.S. Criminal Justice System." Accessed April 20, 2020. https://www.sentencingproject.org/publications/un-report-on-racial-disparities/.

Richardson, Rashida, Jason Schultz, and Kate Crawford. "Dirty Data, Bad Predictions: How Civil Rights Violations Impact Police Data, Predictive Policing Systems, and Justice." SSRN Scholarly Paper. Rochester, NY: Social Science Research Network, February 13, 2019. https://papers.ssrn.com/abstract=3333423.

Rudin, Cynthia, and Joanna Radin. "Why Are We Using Black Box Models in AI When We Don't Need To? A Lesson From An Explainable AI Competition." *Harvard Data Science Review* 1, no. 2 (November 1, 2019). https://doi.org/10.1162/99608f92.5a8a3a3d.

Saunders, Jessica, Priscillia Hunt, and John S. Hollywood. "Predictions Put into Practice: A Quasi-Experimental Evaluation of Chicago's Predictive Policing Pilot." *Journal of Experimental Criminology* 12, no. 3 (September 1, 2016): 347–71. https://doi.org/10.1007/s11292-016-9272-0.

Selbst, Andrew D., Danah Boyd, Sorelle Friedler, Suresh Venkatasubramanian, and Janet Vertesi. "Fairness and Abstraction in Sociotechnical Systems." SSRN Scholarly Paper. Rochester, NY: Social Science Research Network, August 23, 2018. https://papers.ssrn.com/abstract=3265913.

Starr, Sonja B. "Evidence-Based Sentencing and the Scientific Rationalization of Discrimination." SSRN Scholarly Paper. Rochester, NY: Social Science Research Network, September 1, 2013. https://papers.ssrn.com/abstract=2318940.

Stevenson, Megan T. "Assessing Risk Assessment in Action." SSRN Scholarly Paper. Rochester, NY: Social Science Research Network, 2018. https://doi.org/10.2139/ssrn.3016088.

Vagins, Deborah J., and Jesselyn McCurdy. "Cracks in the System: 20 Years of the Unjust Federal Crack Cocaine Law." American Civil Liberties Union. Accessed April 23, 2020. https://www.aclu.org/other/cracks-system-20-years-unjust-federal-crack-cocaine-law.

Van Brunt, Alexa, and Locke Bowman. "Toward a Just Model of Pretrial Release: A History of Bail Reform and a Prescription for What's Next." *Journal of Criminal Law and Criminology* 108, no. 4 (January 1, 2018): 701.

Wexler, Rebecca. "Defendants Should Have the Right to Inspect the Software Code Used to Convict Them." Slate Magazine, October 6, 2015. https://slate.com/technology/2015/10/defendants-should-be-able-to-inspect-software-code-used-in-forensics.html.

Wingfield, Nick. "Amazon Pushes Facial Recognition to Police. Critics See Surveillance Risk." *The New York Times*, May 22, 2018, sec. Technology. https://www.nytimes.com/2018/05/22/technology/amazon-facial-recognition.html.